\documentclass[sigconf]{acmart}
\usepackage{graphicx}
\usepackage{listings}
\usepackage{color} 
\usepackage{threeparttable}
\usepackage{booktabs}
\usepackage{amsmath} 
\usepackage{algorithm} 
\usepackage{algpseudocode} 
\usepackage{natbib}
\usepackage{titlesec}

\usepackage{hyperref}

\AtBeginDocument{%
  }

\copyrightyear{2024}
\acmYear{2024}
\setcopyright{acmlicensed}\acmConference[WWW '24 Companion]{Companion Proceedings of the ACM Web Conference 2024}{May 13-- 17, 2024}{Singapore, Singapore}
\acmBooktitle{Companion Proceedings of the ACM Web Conference 2024 (WWW '24 Companion),
May 13-- 17, 2024, Singapore, Singapore}
\acmDOI{10. 1145/3589335.3651580}
\acmISBN{979-8-4007-0172-6/24/05}

\begin{document}

\title{Unveiling Wash Trading in Popular NFT Markets}

\author{Yuanzheng Niu}
\affiliation{%
  \institution{Hainan University}
  \city{Haikou}
  \country{China}}
\email{niunicole1@outlook.com}

\author{Xiaoqi Li}
\authornote{Corresponding author}
\affiliation{%
  \institution{Hainan University}
  \city{Haikou}
  \country{China}}
\email{csxqli@gmail.com}

\author{Hongli Peng}
\affiliation{%
  \institution{Hainan University}
  \city{Haikou}
  \country{China}}
\email{penghongli2020@outlook.com}

\author{Wenkai Li}
\affiliation{%
  \institution{Hainan University}
  \city{Haikou}
  \country{China}}
\email{liwenkai871@gmail.com}
\thispagestyle{plain}

\begin{abstract}
As emerging digital assets, NFTs are susceptible to anomalous trading behaviors due to the lack of stringent regulatory mechanisms, potentially causing economic losses. In this paper, we conduct the first systematic analysis of four non-fungible tokens (NFT) markets. Specifically, we analyze more than 25 million transactions within these markets, to explore the evolution of wash trade activities. Furthermore, we propose a heuristic algorithm that integrates the network characteristics of transactions with behavioral analysis, to detect wash trading activities in NFT markets. Our findings indicate that NFT markets with incentivized structures exhibit higher proportions of wash trading volume compared to those without incentives. Notably, the LooksRare and X2Y2 markets are detected with wash trading volume proportions as high as 94.5\% and 84.2\%, respectively.
\end{abstract}



\renewcommand{\keywordsname}{\MakeUppercase{Keywords}}
\maketitle

\section{\MakeUppercase{Introduction}}
Blockchain 1.0 introduces the concept of a distributed ledger, which is primarily utilized in cryptocurrencies such as Bitcoin and Litecoin to enable secure transactions and accounting\cite{li2023overview}\cite{1}. Blockchain 2.0, exemplified by Ethereum\cite{buterin2014next}, introduces the paradigm of Turing-complete smart contracts\cite{mao2024automated}, which are self-executing agreements composed of code stored on the blockchain. Non-fungible Tokens (NFTs) \cite{4}, issued on the Ethereum blockchain\cite{zhang2022authros}, represent non-replicable digital assets or unique identifiers managed on the blockchain. They are used to allocate, link, or prove ownership of distinct physical and digital goods.

With the rapid growth of the NFT market, transaction security is facing numerous risks\cite{li2020characterizing}, including market manipulations such as wash trading. Individuals and entities exploit regulatory weaknesses and technical vulnerabilities, attempting to evade oversight measures and disrupt the normal functioning of the NFT market. Uninformed users may incur financial losses by purchasing high-volume NFT assets due to insufficient understanding. For instance, wash trading within the LooksRare market accounts for an estimated 95\% of its activity\cite{song2023abnormal}.

Previous research by Das et al.\cite{understanding} indicates that there are efficiency issues when dealing with large-scale transaction data, coupled with a lack of analysis on the transaction network. Although techniques like graph traversal and the Depth-First Search (DFS) algorithm have been applied to identify fraudulent transactions\cite{von2022nft}, they do not fully address the intricate patterns within vast networks.
In response to these challenges, we propose a novel heuristic algorithm, considering a more comprehensive scenario for detecting wash trade activities within the transaction network. This paper aims to conduct a large-scale data comparison and analysis of wash trading activities by contrasting behaviors across four distinct NFT markets. We collect historical data from four different NFT markets in recent years. Notably, markets like LooksRare and X2Y2 employ incentive mechanisms to encourage user activity in their markets. The main contributions of this paper are as follows:

(1) To the best of our knowledge, we present the first systematic comparative analysis of four popular NFT markets, with specific attention to how changes in market incentive mechanisms impact wash trade activities.

(2) We gather and organize large-scale historical transaction data from four NFT markets, encompassing a total of 25,146,213 transactions. We also develop a heuristic algorithm to effectively detect wash trade activities on NFT markets.

(3) We open-source related codes and datasets at \url{https://figshare.com/articles/dataset/25067936}.

The structure of this paper is as follows. Section 2 offers an introduction to background knowledge and related work. Section 3 encompasses data collection and model design, presenting an abstract model that describes the wash trading process. Section 4 presents experimental findings, and Section 5 concludes our work.
\thispagestyle{plain}

\section{\MakeUppercase{Background}}
\subsection{Wash Trading}
In the field of traditional finance\cite{li2021clue}, wash trading is identified as a manipulative practice in that traders deliberately coordinate transactions to artificially inflate asset prices and trading volumes. This tactic fosters a deceptive impression of a dynamic market environment. Regulatory efforts to curb wash trading trace back to 1936 with the "Commodity Exchange Act," which is characterized as "trading that does not involve actual market risk or change in market position, but is executed with the intent to create a false impression of active buying or selling." This practice is commonly referred to as "round-trip trading"\cite{victor2021detecting}.

\subsection{Trading Rewards}
\begin{figure}[ht]
  \centering
  \vspace{-6pt}
  \includegraphics[width=\linewidth]{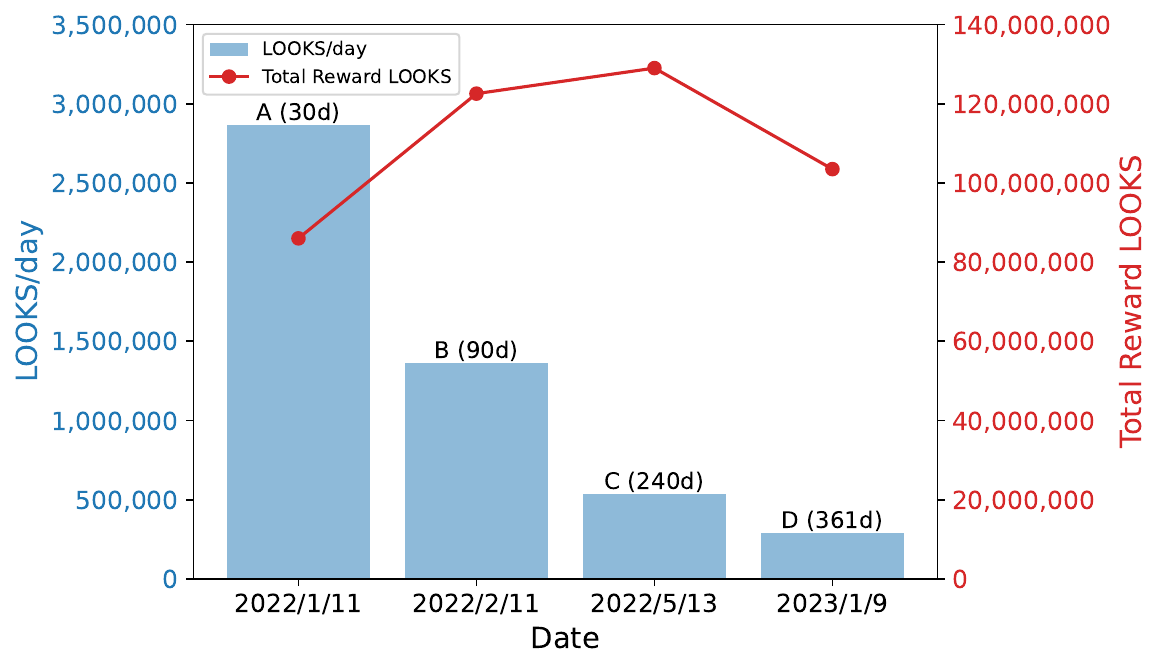}
  \caption{LooksRare Transaction Rewards Allocation Diagram}  
   \vspace{-12pt}
  \label{fig1}
\end{figure}

In certain NFT markets, such as X2Y2 and LooksRare, reward structures incentivize traders with market tokens based on their trading engagement. These tokens, distributed daily and tied to the market's total daily trading volume, offer users opportunities to earn rewards by participating in eligible NFT collectible series transactions on markets\cite{looksrare-rewards}. User rewards are computed by their trading volume ratio to the market's total volume, multiplied by the daily LOOKS token rewards\cite{49}. Consequently, higher trading volumes yield greater rewards, as depicted in Figure \ref{fig1}, where LooksRare's rewards unfold across four phases spanning 721 days.

While both LooksRare and X2Y2 markets have implemented incentive mechanisms, their approaches exhibit significant differences. LooksRare's transaction mining rewards undergo phased reductions, as shown in Figure \ref{fig1}, which potentially influences both trading activity and token prices. In contrast, X2Y2 employs a fixed daily reward distribution strategy, aimed at mitigating market volatility, as noted in \cite{X2Y2-rewards}. Additionally, X2Y2 adopts a cost-effective strategy to attract users by setting its transaction fee at 0.5\%, which is lower than OpenSea's 2.5\% and LooksRare's 2\%. This lower fee structure enhances X2Y2's competitiveness in the market, potentially influencing user preferences and market usage.

\subsection{\MakeUppercase{Related Work}}
Various methods, such as machine learning and graph analysis, have emerged in the field of digital asset analysis for analyzing wash trades.
Victor et al. \cite{victor2021detecting} conduct the inaugural systematic analysis and detection of wash trading on decentralized exchanges IDEX and EtherDelta. They not only quantify the financial scope of these activities but also pinpoint the specific accounts and transaction structures implicated in such trading. Based on on-chain data, Cui et al. \cite{cui2023wteye} devise a system for detecting ERC20 token wash trades, utilizing algorithms to identify and quantify trading patterns. Serneels\cite{serneels2023detecting} explores wash trade detection from three analytical angles, including closed-loop detection and transaction volume analysis. Utilizing data mining and machine learning, Son et al. \cite{song2023abnormal} delve into the identification of fraudulent activities within the NFT market. Concurrently,  Wen et al\cite{wen2023nftdisk} develop NFTDisk, an innovative visualization tool that assists in detecting wash trades in the NFT market, featuring a Disk Module for address reordering and visual clutter reduction and a Flow Module for multi-level transaction visualization.
Chen et al. \cite{chen2023dark} conduct a multidimensional analysis of NFT wash trading, examining its impact on the market and user behavior. This paper conducts a comparative analysis of four distinct NFT trading markets, examining how changes in market incentive mechanisms affect wash trading activities.
\thispagestyle{plain}
\section{\MakeUppercase{Data collection And Study design}}
\subsection{Data Collection}
We utilize Infura\cite{infura} as the access point to connect to the Ethereum network and retrieve raw data via the provided API. This includes block information, transaction details, and event logs for smart contracts. Following the acquisition of blockchain data, we process the information. Customized data parsing scripts are designed for each market, enabling the identification and extraction of NFT transaction-relevant information from the blockchain data. Subsequently, the data are standardized, encompassing block number, timestamp, transaction hash, participants, collection, tokenId, and price. Finally, timestamps are converted to time, and the price data is transformed into corresponding amounts.

\subsection{Abstraction Model for Wash Trading }

\begin{figure}[!ht]
  \centering
   \vspace{-8pt}
  \includegraphics[width=\linewidth]{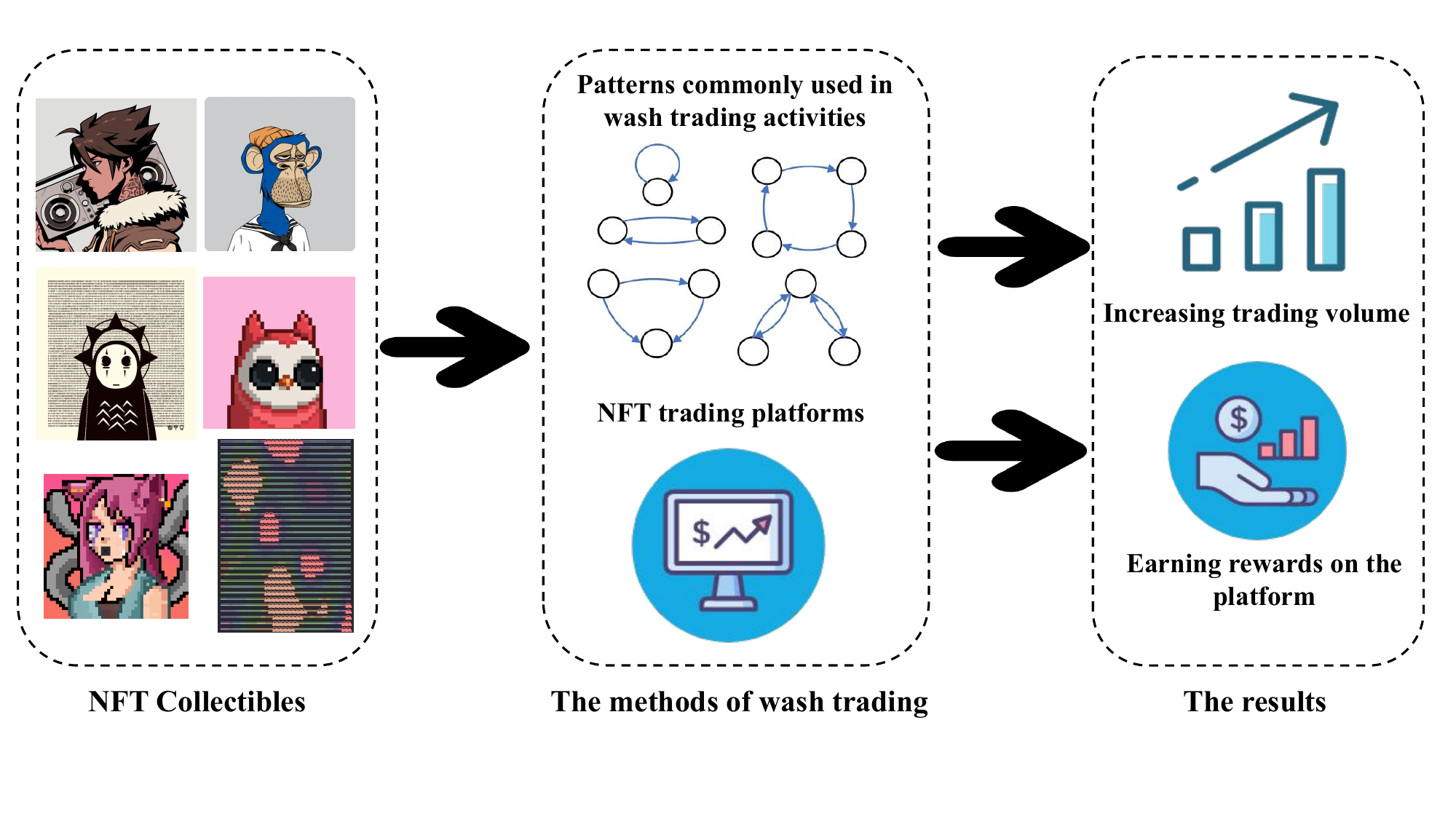}
  \vspace{-22pt}
  \caption{Illustration of the Wash Trading Phases for NFTs}
  \vspace{-6pt}
  \label{fig2}
\end{figure} 

Figure \ref{fig2} illustrates the process of wash trading among NFT collectibles in the NFT market. In particular, wash traders initiate the process by purchasing NFT collectibles on various major NFT markets. Following successful acquisitions, wash traders engage in wash trading through various methods in a repetitive cycle, ultimately aiming to boost trading volumes and earn rewards.
\vspace{-2ex}
\begin{algorithm}
\caption{Wash Trading Detection Algorithm}
\label{al1}
\begin{flushleft}
\textbf{Data:} "transactions.csv", parameters $\beta$, $\psi$, $\omega$ \\
\textbf{Input:} $"pre_{params}", "G_{params}" $\\
\textbf{Result:} "$topo_{features}$ ", "$suspect_{nodes}$", "$suspect_{trans}$"
\end{flushleft}
\begin{algorithmic}[1]
\State $data, G \gets create\_graph("transactions.csv", pre_{params}, G_{params})$
\State $topo, suspect_{nodes}, suspect_{trans} ,G_{census} \gets analysis(G) $
\State $washtrades \gets get\_trans\_scc(G)$
\State $washtrades\_V \gets get\_trans\_V(G)$
\For{each node in $G$}
    \State $node\_trans \gets get\_transactions(node)$
    \If{is\_suspicious($node\_trans$, $topo$, $\beta$)}
        \If{transaction\_count($node\_trans$) $> \omega$}
            \State $suspect_{nodes}[n] \gets ``in\_node''$
            \State $remove_{node}(G, node)$
        \Else
            \State $node\_trans \gets filter\_trans(node\_trans, topo)$
            \State $label\_node(node, ``wash\_node'')$
            \State $n\_xt \gets getunmarkedneighbor(node, G)$
            \State $service\_n \gets get\_addr(node\_trans)$
            \State $suspect_{trans}[service\_n] \gets ``service\_n''$
        \EndIf
        \State $suspect_{trans}[n] \gets node\_trans$
    \ElsIf{belongs\_to\_scc(node, washtrades)}
        \State $label\_node(node, ``scc\_wash\_node'')$
    \ElsIf{part\_of\_V(node\_trans, washtrades\_V)}
        \State $label\_node(node, ``V\_wash\_node'')$
    \EndIf
\EndFor
\State $summarize\_results(topo, suspect_{nodes}, suspect_{trans})$
\end{algorithmic}
\end{algorithm}
\vspace{-2ex}
Wash trading typically manifests as cyclic or repetitive trades within the network, with such transactions being more prominent within the strongly connected components of the network.
Algorithm \ref{al1}, Wash Trading Detection Algorithm, adopts a graph analysis methodology to unearth such illicit activities by examining the balance drift of transactions within these components.

The process extracts transaction records from a "transactions.csv" file. The algorithm then sets threshold parameters "$\beta$", "$\psi$", and "$\omega$" to tune the sensitivity of the detection phase. During the data preprocessing stage, it constructs a transaction graph utilizing the parameters "$pre_{params}$" and "$G_{params}$".
Topological features, denoted by "$topo_{features}$", are initialized to uncover potential wash trading within the network. The algorithm also logs nodes that raise suspicion, labeled as "$suspect_{nodes}$", alongside their corresponding transactions, "$suspect_{trans}$". The detection strategy involves scrutinizing transactions for any signs of suspicion as governed by the "$\beta$" parameter, with subsequent node filtering based on "$\psi$" and "$\omega$".

Upon preprocessing the data, which includes reading the transaction data and generating the graph (shown in step 1 of Algorithm \ref{al1}),  it subsequently processes and filters transactions, particularly those that involve repurchases (shown in steps 5-24 of Algorithm \ref{al1}). The algorithm employs two principal methods to detect wash trading behaviors: firstly, by calculating the average number of transactions in strongly connected components, and secondly, by assessing the percentages of transaction volumes (shown in steps 3-4 of Algorithm \ref{al1}).
The algorithm analyzes the transaction graph's structure, focusing on binary and ternary relationships, to expose wash trade patterns (as part of the 'analysis' function shown in steps 1-2 of Algorithm \ref{al1}). It then iteratively traverses each node in the transaction graph, assessing if the associated transactions appear dubious for every node using the is\_suspicious function. If a node is deemed suspicious and the transaction count surpasses the "$\omega$" threshold, it is marked as a participant in wash trading activities (shown in steps 8-10 of Algorithm \ref{al1}). Conversely, if the node does not meet these criteria, it is simply labeled, and the algorithm continues to track the related service addresses (shown in steps 12-16 of Algorithm \ref{al1}).

\section{RESULT}
\begin{table*}[h]
\centering
\caption{Wash Trading Detection Results in Popular NFT Markets}
\vspace{-2ex}
\label{table 1}
  \begin{threeparttable} 
\begin{tabular}{cccccccc}
\toprule
{} & {} & \multicolumn{3}{c}{\textbf{Number of Transaction}} &\multicolumn{3}{c}{\textbf{Volume of Transaction (ETH)}}\\
       \cmidrule(rl){3-5} \cmidrule(rl){6-8} 
\textbf{Market} & \textbf{Rewards} & \textbf{Total} &\textbf{Wash Trades} &\textbf{\% of Wash Trades} & \textbf{Total} &\textbf{Wash Trades} &\textbf{\% of Wash Trades}\\
\midrule
    OpenSea & $\times$ & 22,072,332 & 123,955 &0.5\%  & 37,236,607,173.1 & 9,330,199,299.4 & 25.1\%  \\
    
    Blur & $\times$  & 924,386 &  83,436 & 9.0\%  &  743,557.1 & 222,395.5 &  29.9\% \\
    
LooksRare & $\checkmark$ & 386,100  &  85,288 & 22.1\% & 10,126,545.8 &  9,575,596.6 & 94.5\% \\
     
     X2Y2 & $\checkmark$ & 1,763,395 & 364,223 & 20.6\% & 2,724,651.1 & 2,296,475.5 & 84.2\% \\
    \bottomrule
   \vspace{-5ex}
    \end{tabular}%
    \end{threeparttable} 
\end{table*}%

\begin{figure}[!ht]
  \centering
  \vspace{-2ex}
  \includegraphics[width=1\linewidth]{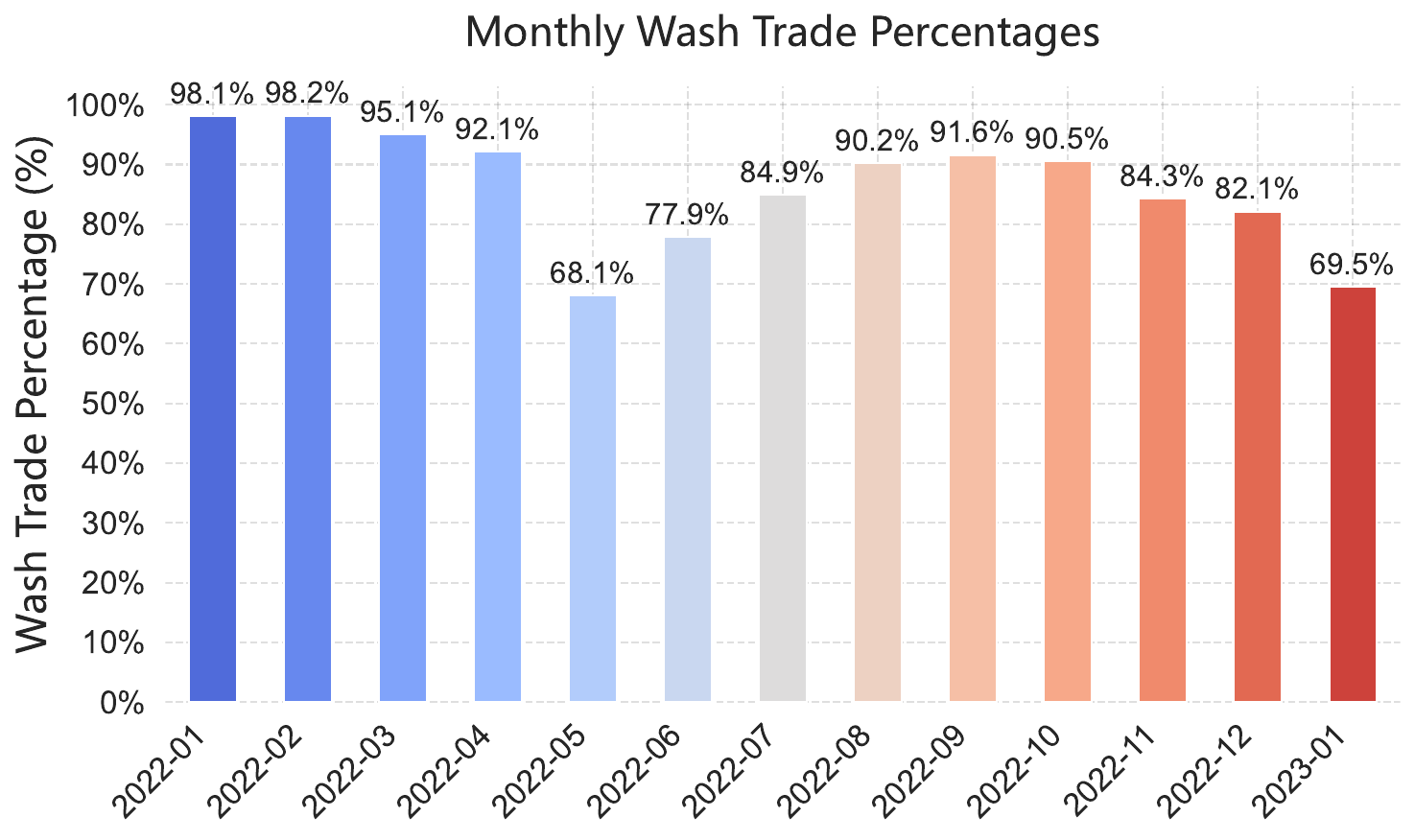} 
  \vspace{-4ex}
  \caption{Wash Trading Volume in LooksRare}
  \vspace{-2ex}
  \label{fig3}
\end{figure}

As illustrated in Figures \ref{fig1} and \ref{fig3}, the volume of wash trading in the LooksRare market exhibits notable fluctuations in response to changes in its reward mechanisms. With a reduction from a peak of 98.2\% to 69.5\% between January 2022 and January 2023, as depicted in Figure \ref{fig3}. The reward distribution mechanism of the LooksRare market, highlighted in Figure \ref{fig1}, demonstrates that while the daily allocation of LOOKS tokens decreases across Phases A, B, and C, the cumulative volume of "Total Reward LOOKS" witnesses an increase. During Phases A and B, from January to April 2022, wash trade volumes consistently remain above 92\%. Entering Phase C (May 2022), a substantial reduction in daily LOOKS rewards occurs, and the overall growth of "Total Reward LOOKS" begins to plateau, which coincides with a considerable decrease in wash trade proportions. In Phase D, characterized by a decline in both the total quantity of rewards and daily issuance of LOOKS tokens, there is a significant decrease in wash trading on the LooksRare market.

\thispagestyle{plain}

This reward system coincides with a considerable decrease in wash trade proportions, inadvertently fostering and amplifying wash trading activities. This results in artificially inflated trade volumes that do not accurately reflect the true dynamics of the market. Concurrently, LooksRare confronts challenges arising from the diminishing efficacy of its reward mechanisms.

\begin{figure}[!ht]
  \centering
 \vspace{-4ex}
  \includegraphics[width=\linewidth]{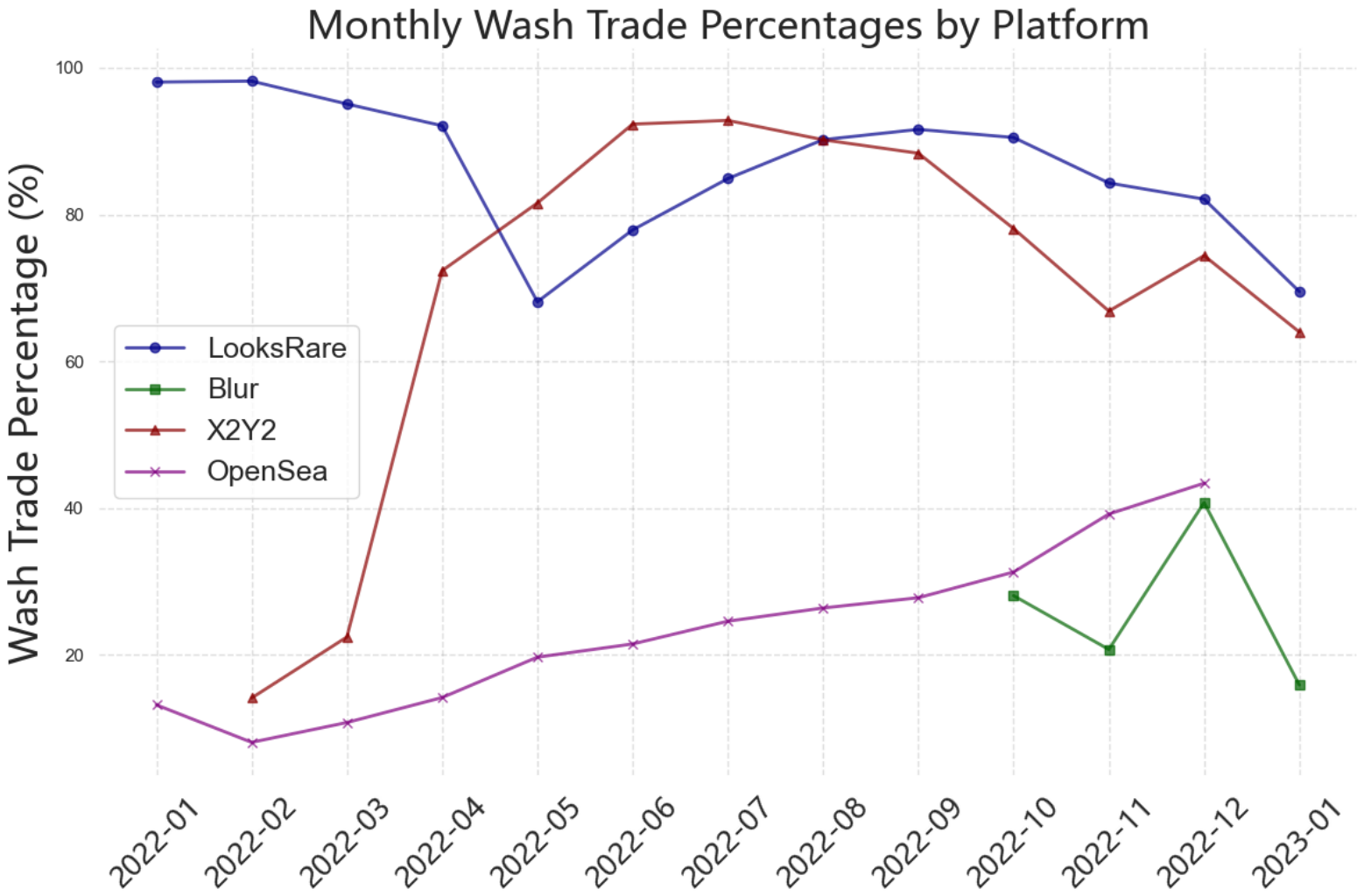}
   \vspace{-6ex}
  \caption{Comparative Analysis of Wash Trading Volumes Across Four NFT Markets}
  \label{fig4}
  \vspace{-2ex}
\end{figure}

Figure \ref{fig4} presents a graphical analysis of wash trading volumes across four NFT markets from 2022 to 2023. Moreover, compared to other NFT markets, LooksRare and X2Y2 continue to exhibit high wash trade ratios under their incentive policies, suggesting that unique market rewards may facilitate or tolerate such behaviors. In our analysis of wash trading within various NFT markets, Table \ref{table 1} indicates that markets with reward mechanisms exhibit significantly higher numbers and volumes of wash trades. Specifically, LooksRare and X2Y2, both of which offer rewards for trading, have wash trade percentages of 22.1\% and 20.6\% respectively, in terms of transaction count. The volume of these wash trades is more striking, representing 94.5\% and 84.2\% of the total ETH volume on each market. The data indicate that reducing incentives can decrease wash trade occurrences, yet addressing the issue more effectively may require intricate market regulatory mechanisms and enhanced transparency. Additionally, our analysis underscores the need for in-depth research into the incentive structures and market behaviors of different markets to develop more nuanced regulatory frameworks.

\section{CONCLUSION}
We present an analysis of wash trading behaviors across four major NFT markets, particularly highlighting how the rules set by different markets affect wash trading actions. We introduce an algorithm that uses graph theory to build a trading network graph for detecting wash trading. Our method analyzes strongly connected components and binary and ternary relationships within the network to identify cyclic trading patterns among nodes. The findings show a marked contrast in wash trading activities: minimal on OpenSea and Blur, yet significantly higher on LooksRare and X2Y2, with volumes of 94.5\% and 84.2\%, respectively. This disparity underscores the impact of reward mechanisms as potential catalysts for wash trading.
Furthermore, the study notes that although reward mechanisms may draw users and volume to LooksRare and X2Y2 in the short term, they pose a challenge to the long-term sustainability of these markets. The influx of users engaging in wash trading to artificially inflate NFT prices undermines market liquidity, poses a risk of user attrition, and leads to declining staking rewards, with the low cost of switching markets further contributing to reduced trading rewards. This cycle potentially results in a fall in token prices. We plan to utilize real-time market data and user behavior patterns to detect wash trading. 

\vspace{-1ex}
\bibliographystyle{plainnat} 
\bibliography{main}
\thispagestyle{plain}
\end{document}